\documentclass[
    twocolumn,
	prl,
	amssymb,
	preprintnumbers, superscriptaddress,
	nofootinbib]{revtex4-2}
\pdfoutput=1
\usepackage[
pdfauthor={Djuna Croon}]{hyperref}
\hypersetup{
    colorlinks=true,
    linkcolor=red,
    citecolor=blue,
    filecolor=black,
    urlcolor=black,
}

\usepackage{amsmath} 
\usepackage[T1]{fontenc}
\usepackage{bm}
\usepackage{graphicx}
\usepackage{caption}
\usepackage{subcaption}
\usepackage{slashed}
\usepackage[normalem]{ulem}
\usepackage{gensymb}
\usepackage{comment}
\usepackage{lipsum}
\usepackage[table]{xcolor}
\usepackage{hhline,colortbl}
\usepackage{boldline}

\newcommand{\beq}{\begin{equatxion}} 
\newcommand{\eeq}{\end{equation}}  
\newcommand{\bea}{\begin{eqnarray}}  
\newcommand{\eea}{\end{eqnarray}}  
\newcommand{\pd}[2]{\frac{\partial #1}{\partial #2}}
\newcommand{\pdd}[2]{\frac{\partial^2 #1}{\partial #2^2}}

\newcommand{\Refs}[1]{Refs.~\cite{#1}}

\allowdisplaybreaks
\newcommand{\mape}{\bar{E}_{\%}}
\newcommand{\ape}{E_{\%}}

\newcommand{\se}[1]{\begin{align}\begin{split} #1 \end{split}\end{align}}

\renewcommand{\paragraph}[1]{~\\ \noindent{\bf \emph{#1} --}}
\renewcommand{\subparagraph}[1]{~\\ \noindent{ \emph{#1} --}}

\usepackage{natbib}
\usepackage{graphicx}

\allowdisplaybreaks

\newcommand{\pL}{\left(} 
\newcommand{\pR}{\right)}

\begin{document}

\title{Machine learning a manifold}
\author{Sean Craven}
\affiliation{Department of Physics, Durham University, Durham DH1 3LE, U.K.}
\author{Djuna Croon}
\email{djuna.l.croon@durham.ac.uk}
\affiliation{Department of Physics, Durham University, Durham DH1 3LE, U.K.}
\affiliation{Institute for Particle Physics Phenomenology, Durham University, Durham DH1 3LE, U.K.}
\affiliation{TRIUMF Theory Group, 4004 Wesbrook Mall, Vancouver, B.C. V6T2A3, Canada}
\author{Daniel Cutting}
\email{daniel.cutting@helsinki.fi}
\affiliation{Department  of  Physics  and  Helsinki  Institute  of  Physics,  PL  64,  FI-00014  University  of  Helsinki,  Finland}
\author{Rachel Houtz}
\email{rachel.houtz@durham.ac.uk}
\affiliation{Department of Physics, Durham University, Durham DH1 3LE, U.K.}
\affiliation{Institute for Particle Physics Phenomenology, Durham University, Durham DH1 3LE, U.K.}

\preprint{IPPP/21/56}

\date{\today}
\begin{abstract}
We propose a simple method to identify a continuous Lie algebra symmetry in a dataset through regression by an artificial neural network. 
Our proposal takes advantage of the $ \mathcal{O}(\epsilon^2)$ scaling of the output variable under infinitesimal symmetry transformations on the input variables. 
As symmetry transformations are generated post-training, the methodology does not rely on sampling of the full representation space or binning of the dataset, and the possibility of false identification is minimised.
We demonstrate our method in the SU(3)-symmetric (non-) linear $\Sigma$ model. 
\end{abstract}
\maketitle

\paragraph{Introduction}
Symmetry principles have drastically simplified the description of particle physics in the twentieth century. 
Famously, the 8-fold way \cite{Gell-Mann:1962yej} of organizing pions and kaons into a representation of an
approximate SU$(3)$ flavor symmetry lead to the development of the quark model. In the same vein, 
future discovery experiments would primarily have access to the low-energy particle content of theories beyond the standard model (BSM): in the case of a broken approximate global symmetry, this includes the pseudo-Nambu Goldstone bosons (pNGB), transforming under the adjoint representation of the unbroken symmetry (see e.g. ~\cite{Low:2014nga} for a relevant review.).
If the BSM theory is confining, the symmetries of the low energy theory provide a window to the structure of the high energy theory through the barrier of the strong coupling regime. 
However, the pNGB
representation need not have a small dimensionality, or define a simple topology. It may also be broken both spontaneously and explicitly, and the dataset may be noisy.
Identifying residual (approximate) symmetries is therefore an interesting problem. 

Motivated by this problem, we investigate the use of artificial neural networks (NN) to identify a symmetry in a dataset. 
We work with a simplified version of the problem: a function $ V(\phi)$ symmetric under a transformation of coordinates $\phi \to f(\phi)$: $ V(\phi) = V(f(\phi))$. To interpolate between datapoints we use a NN
(recently discussed in the context of high energy physics in \cite{Chahrour:2021eiv}), which allows
us to test the local properties of the manifold and deduce the presence of a symmetry -- or rather, eliminating the possibility of its absence -- from its topology. 

Detection of symmetry with the use of machine learning has a long history~\cite{Sejnowski1986LearningSG}, 
though most
attempts focus on mirror or rotational symmetries in image data and within the domain of computer vision~\cite{Konen2008UnsupervisedSD,Tsogkas2012LearningBasedSD,Nagar20203DSymmRA,George2021SymmetryPW}. In recent years there has been an increased interest in learning invariant transformations of input data which do not change the output of a specific machine learning task~\cite{Cohen2017SteerableC,vanderWilk2018LearningIU,benton2020learning,Dehmamy2021AutomaticSD,Romero2021LearningEA,Mourdoukoutas2021ABA,zhou2021metalearning,Dillon:2021gag}. 
This is useful as the construction of invariant or equivariant NN reduces the number of samples of input data required for generalization. 
Machine learning has been used to explore various features of conformal field theories, including to distinguish between scale invariant and conformal symmetries~\cite{Chen:2020dxg}. It has also been demonstrated that computation of tensor products and branching rules of irreducible representations are
machine-learnable~\cite{Chen:2020jjw}.
Furthermore, a recent work has investigated using generative adversarial networks to learn transformations that preserve the measured probability density function of a random process~\cite{desai2021symmetrygan}. 
Here we are interested in a variation to this problem, testing for the presence of a symmetry in a dataset that samples a patch of a function whose domain has a high dimensionality.

The use of NN for the detection of symmetries in such a context has previously been considered by \cite{Krippendorf:2020gny, Barenboim:2021vzh} for translations, discrete symmetries, and  SO$(N) \simeq$ SU$(N-1)$ with $N< 3$. 
The methodology in this paper differs from the approaches taken in \Refs{Krippendorf:2020gny, Barenboim:2021vzh} in two important aspects. 
Firstly, 
points related by a symmetry transformation are generated post-training. 
This implies that the local properties of the manifold can in principle be studied without global knowledge of the manifold, or a large number of close neighbors in the tangent space. Both of these implications may prove to be
a marked advantage in datasets with large dimensionality. For example, no pre-training stage of narrow bin definition and data categorization is necessary. The fraction of data which is related by the action of a single generator roughly scales like $1-(\Delta y/y)^d $, where $\Delta y$ is a narrow bin width in output variable $y$ and where $d$ is the dimensionality of the dataset. 
A further difference is that 
the methodology here can be used to demonstrate the absence of a symmetry, such that the probability of mis-identification is minimized.
With sparser sampling
the assumptions made about the symmetry transformation (for example its direction)
may play an increasingly important role, potentially leading to the false identification of SO$(N)$ symmetry. 
We demonstrate in particular that our methodology can be used to show the absence of SO$(8)$ (and the presence of SU$(3)$) using the non-linear sigma model.

\paragraph{Methodology}
To detect the Lie algebra, we take advantage of the fact that the symmetry is continuous and locally defined. In the presence of a symmetry, an infinitesimal transformation of the fields of the form $ \phi_i \to \phi_i'= \phi_i + \epsilon T_{ij} \phi_j$ leads to a transformation of the effective action of $\mathcal{O}(\epsilon^2)$:
    \se{
    V   &\to V +
    \epsilon\pL \pd{V}{\phi_i}\delta \phi_i + \pd{V}{(\partial_\mu \phi_i)} \delta \partial_\mu \phi_i \pR
    +\mathcal{O}(\epsilon^2)
    \\
    \Rightarrow
    V   &\to V + \mathcal{O}(\epsilon^2)\ .
    }
The $\mathcal{O}(\epsilon^2)$ and higher terms remain as the Lie algebra lives in the tangent space of the Lie group's manifold.
They are (for simplicity focusing on a single multiplet without derivative interactions):
\se{
V   &\to V + \epsilon^2  \pdd{V}{\phi_i} (\delta \phi_i)^2   + \mathcal{O}(\epsilon^3)
}

A neural network can be used to interpolate a dataset and make predictions for the transformed fields. Then, if the symmetry is present, we should find
\se{ 
\label{eq:DeltaV}
(\Delta V)_{\rm NN} 
&\equiv \left| \frac{V_{\rm NN}(\phi_i')- V(\phi_i)}{V(\phi_i)} \right| \\
&= \epsilon^2 \, \frac{V_{\rm NN}''(\phi_i)}{V(\phi_i)} 
(\delta \phi_j)^2  
+ \frac{\ape}{100 \%}  + \mathcal{O}(\epsilon^3) \ ,
}
where  \se{\ape &= \left|\frac{V_{\rm NN}(\phi_i)- V(\phi_i)}{V(\phi_i)} \right|\times 100 \%} is the absolute percentage error of the NN on the validation set, $\phi$ is a datapoint in the validation set, $\phi'$ its image under the transformation to be tested, and $ V_{\rm NN}(\phi_i')$ is the NN prediction of the transformed field.
As $\phi_i$ is part of the dataset, $V(\phi_i)$ is known and does not need to be predicted by the network. 
Importantly, the $ \epsilon^2$ scaling is independent of the normalization, and $ V_{\rm NN}''(\phi_i)(T_{ij} \phi_j)^2/V(\phi_i) \sim 1$ approximates the a-priori unknown coefficients in the expansion.

\paragraph{Models}
We use as inspiration a BSM scenario of a new scale of spontaneous symmetry breaking (SSB) that leaves behind Nambu Goldstone Boson (NGB) fields.\footnote{In a realistic model, explicit breaking of the nonlinearly realized symmetry would lift the NGB masses, and these would actually be pNGBs. We leave a study of explicit breaking to future work, and will refer to these fields as NGBs. Note that the linearly realized flavor symmetry studied here could remain preserved under such explicit breaking.}
These would generically be the lightest fields and a reasonable guess as the earliest indication of the new physics. The symmetries exhibited by NGB interactions would then be a probe of the structure of the theory at or above the symmetry breaking scale.  

The interactions of the NGBs are parameterized by low-energy effective theories of spontaneously broken symmetries. We will focus on two such benchmark models, the linear and non-linear $\Sigma$ models.

\subparagraph{Non-linear  $\Sigma$  model}
The non-linear $\Sigma$ model (NL$\Sigma$M) is given by:
\se{
\mathcal{L}
&= \frac{ f^2 }4 \text{tr}\left( \partial_\mu \Sigma  \partial^\mu \Sigma^\dagger \right) \ ,
\quad
\Sigma   = \exp\left( i   \pi^a t^a / f \right) \ .
}
which has a non-linearly realized $SU(N)_L \times SU(N)_R$ chiral symmetry and a preserved $SU(N)_F$ flavor symmetry below the SSB scale $f$.

As the flavor symmetry is manifest order by order in $f$, we can 
expand to $\mathcal{O}(1/f^2)$ to obtain:
    \se{
    V
        =& 
        - \frac1{24 f^2} \left[ \partial_\mu \pi^a \pi^b \partial^\mu \pi^c \pi^d 
            - \pi^a \partial_\mu \pi^b \partial^\mu \pi^c \pi^d \right] \\ & \times \text{tr}\left( t^a t^b t^c t^d \right) \ .
    \label{eq:NLsigma}
    }
The pions of~(\ref{eq:NLsigma}) are in the adjoint representation of $SU(N)_F$, and transform as
    \se{
    \pi^a
        &\to \pi^a + \epsilon f^{abc}  \Theta^b \pi^c\ ,
    \label{eq:suN-pi}
    }
where $\epsilon \Theta^a$ gives a set of infinitesimal transformation parameters. The $f^{abc}$ are the structure constants of $SU(N)$
    \se{
    [t^a, t^b] = i f^{abc} t^c \ ,
    } 
which form the Lie algebra. Under the transformation in~\eqref{eq:suN-pi}, the potential changes as $V \to V + \mathcal{O}(\epsilon^2)$.

Note that for $N\leq2$, $SU(N) \simeq SO(N-1)$. Our goal is to identify the  $SU(N)$ flavor symmetry of the NL$\Sigma$M, and in general $SU(N)$ will not be isomorphic to any $SO(N)$ group. Consider the NL$\Sigma$M with the the lowest $SU(N>2)$ flavor symmetry. In this case the pions form an $8$-plet in the adjoint representation of $SU(3)$, but could also be rotated under $SO(8)$. Acting an $SO(8)$ transformation on these pions gives
    \se{
    \pi_i
        &\to  R_{ij} \pi_j
        = \pi_i + \epsilon {T^a_{ij}} \Theta^a \pi_j + \mathcal{O}(\epsilon^2)\ ,
    }
where $T^a$ are the generators of the $SO(8)$ Lie algebra. This yields 
    \se{
    V   &\to V + \mathcal{O}(\epsilon) \ ,
    }
as one would expect for an infinitesimal  transformation not associated with a symmetry of the theory. The ability to disentangle $SU(3)$ from $SO(8)$ is thus required to detect the correct symmetry present in the NL$\Sigma$M.

To summarize, the NL$\Sigma$M with an $N= 3$ flavor symmetry changes as:
    \se{
    V   &\xrightarrow{SU(3)} V + \mathcal{O}(\epsilon^2) \quad \  
        \text{symmetry present}
    \\
    V   &\xrightarrow{SO(8)} V + \mathcal{O}(\epsilon) \quad \quad \!
        \text{symmetry absent}
    }
under $SU(3)$ and $SO(8)$ transformations of the $\pi^a$ fields. This behavior will be exploited in our symmetry detection strategy below.

\subparagraph{Linear $\Sigma$  model}
The same symmetry pattern $SU(N)_L \times SU(N)_R \to SU(N)_F$ can be described by the linear $\Sigma$ model (L$\Sigma$M),  given by 
    \se{
    \mathcal{L} 
        =& \text{tr}\left( \partial_\mu \Sigma \partial^\mu \Sigma^\dagger \right)
            + m_\Sigma^2 \text{tr}\Sigma\Sigma^\dagger
            + \left(  \mu_\Sigma\det\Sigma + h.c. \right)
            \\ & 
            - \frac\lambda2 \left( \text{tr}\Sigma\Sigma^\dagger \right)^2
            - \frac\kappa2 \text{tr}\Sigma\Sigma^\dagger\Sigma\Sigma^\dagger \ ,
   }
where
   \se{
    \Sigma_{ij}
        &= \frac{ \varphi + i \eta' }{ \sqrt{2N} } 
            + X^a t^a_{ij} + i \pi^a t^a_{ij}.
    \label{eq:Lsigma}
    }
Working from the assumption that the NGB fields will be the lightest, we integrate out the heavy $X, \varphi$ fields associated with unbroken generators. For simplicity, we assume sufficient symmetry breaking effects to lift the mass of the $\eta'$ field enough so that it may also be integrated out.\footnote{If~(\ref{eq:Lsigma}) describes the low energy theory behavior of QCD-like confinement of some non-Abelian gauge field, the corresponding $\eta'$ would generically acquire a mass of the order of $m_{X, \varphi}$ due to explicit $U(1)_A$ breaking from instanton effects.} This leaves only the pion field interactions:
    \se{
    \mathcal{L}
        &= \frac12 \partial_\mu \pi^a \partial^\mu \pi^a
            + \frac12 m_\Sigma^2 \pi^a \pi^a
            + \frac\lambda{8} \left(\pi^a \pi^a \right)^2
        \\
        &\quad + \frac\kappa{2}\text{tr}\left( t^a t^b t^c t^d \right) \pi^a \pi^b \pi^c \pi^d\ .
    \label{eq:LSigmaM} }
This potential is again invariant under the $SU(N)$ transformations of~(\ref{eq:suN-pi}).
Unlike the NL$\Sigma$M, however, the potential in \eqref{eq:LSigmaM} is also invariant under an $SO(8)$ symmetry for $N=3$ flavors:
    \se{
    V_\text{L$\Sigma$M}   &\xrightarrow{SU(3)} V_\text{L$\Sigma$M} + \mathcal{O}(\epsilon^2) 
    \\
    V_\text{L$\Sigma$M}   &\xrightarrow{SO(8)} V_\text{L$\Sigma$M} + \mathcal{O}(\epsilon^2)  \ . 
    }
We therefore expect to be able to detect the presence of both symmetries. It will be useful to contrast a symmetry transformation with a non-symmetric transformation. For this purpose we use 
the simple transformation:
    \se{
    \pi_i &\to \pi_i + \epsilon_{ij} \pi_j\ , 
    \ \ \ \epsilon_{ij}= \epsilon \times \frac1{n^2}\ \  \text{for all } i, j \ ,
    \label{eq:faux}
    }
where $n$ is the number of $\pi$ fields. This transformation does not correspond to any symmetry and changes the potential as $V_\text{L$\Sigma$M} \to V_\text{L$\Sigma$M} + \mathcal{O}(\epsilon)$. We will refer to this transformation as arb(8) 
in the rest of this paper.

\begin{table}
\begin{tabular}{|l | l |}
 \hline
 Hyperparameter $\quad$ & Value  \\ [0.5ex] 
 \hline\hline
 Hidden layers & 8 \\ 
 \hline
 Neurons/layer & 512  \\
 \hline
 Optimizer & Adam \\
 \hline 
 Learning Rate & $10^{-3}$,  $\beta_1 = 0.9, \, \beta_2 = 0.99$  \\
 \hline
 Loss function & MAPE \\
 \hline
 Training epochs & 215  \\
 \hline
 Training set size & $10^6 \times 0.9$\\
 \hline
 Batch size & 16 \\
 \hline
\end{tabular}
\caption{\small Neural network hyperparameters
}
\label{tab:nnhypers}
\end{table}

\paragraph{Neural network}
The methodology proposed above uses a sequential feed-forward neural network to perform regression. As a result of the universal approximation theorem (UAT) \cite{Cybenko:1989, Hornik:1991}, there is no theoretical limit to the accuracy with which a neural network with a single hidden layer and enough neurons can approximate any function. 
Moreover, as was recently demonstrated in \cite{Chahrour:2021eiv}, hidden layers can increase the interpolating abilities of the NN (the L$\Sigma$M \eqref{eq:Lsigma} and NL$\Sigma$M \eqref{eq:NLsigma} SU(3) potentials contain 80
and 143 terms from 8 and 16 dimensional input respectively).
In this section we report on neural network's architecture and hyperparameters used in the analysis below. We motivate these choices in the supplementary material. 

To create our neural networks we used the Keras~\cite{chollet2015keras} library. The neural networks used had 8 hidden layers with 512 neurons with hyperparameters as in Table~\ref{tab:nnhypers}, but we observed no strong dependence on this architecture. We found the best performance with an adaptive learning rate activation function with a small initial learning rate. No markers of overtraining were observed. 

The training data was generated using uniform sampling in $ |\phi|^{1/4}$, where $\phi = \{ \pi, \partial_\mu \pi \} $ represents an input field.
This distribution of input points was chosen to get an approximately Gaussian distribution in $ V(\phi)$. We found that the network's $\mape$ performance scaled monotonically with the training set size, as expected. Notably, the performance was inversely correlated with batch size >8, which we attribute to the network effectively averaging out important features of the manifold.

\begin{figure}
    \centering
    \begin{minipage}{.49\textwidth}
    \includegraphics[width=\textwidth]{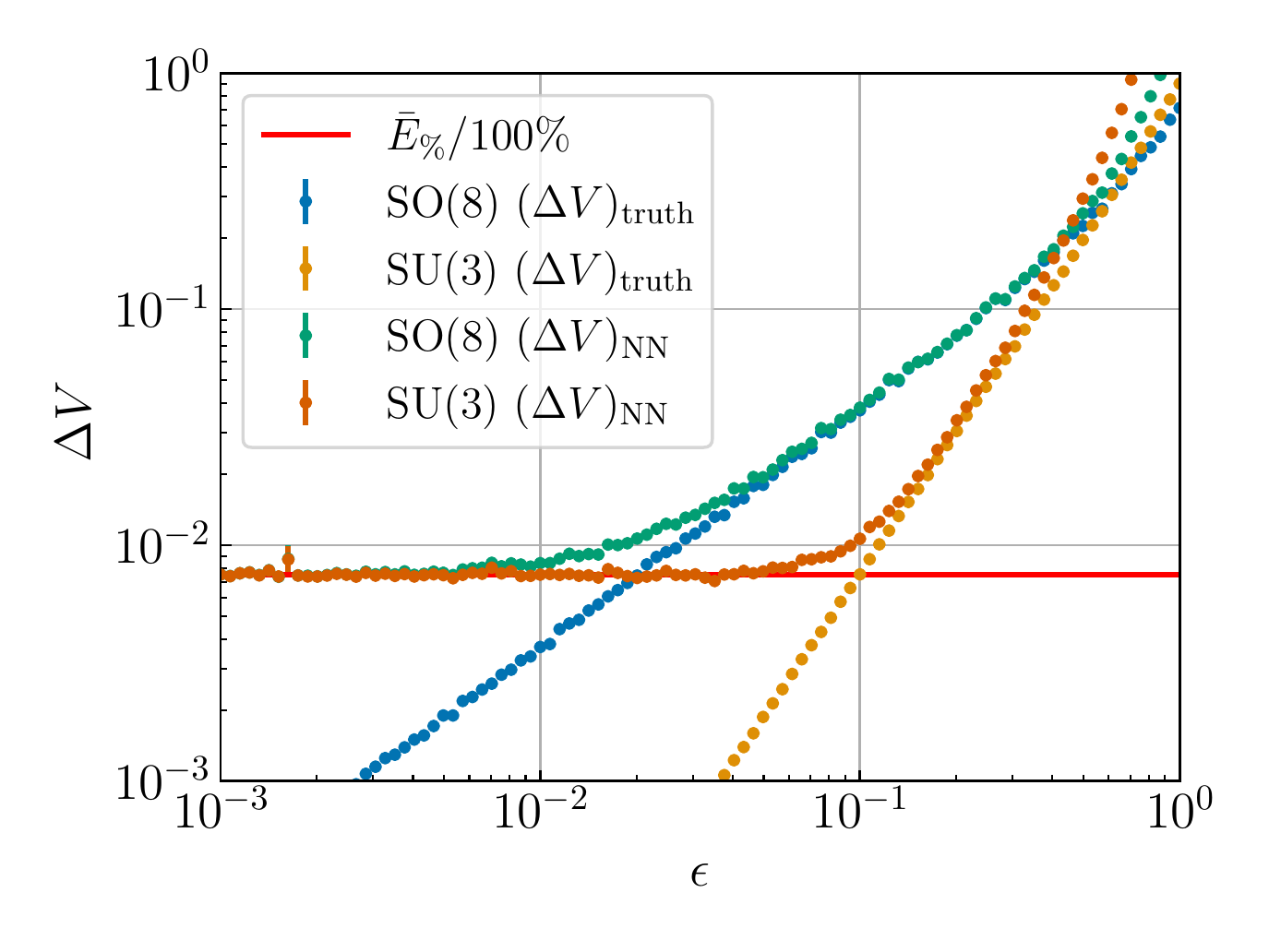}
    \caption{\small The value of 
    $(\Delta V)_{\rm truth}$ or $(\Delta V)_{\rm NN}$ as a function of $\epsilon$ for the NL$\Sigma$M \eqref{eq:NLsigma}.
    The horizontal red line indicates the converged $ \mape/100\%$, which corresponds to the expected noise floor for $(\Delta V)_{\rm NN}$. Near $\epsilon \to 1$ higher order terms in the expansion \eqref{eq:DeltaV} become important. Error bars on the data points correspond to the standard error on the mean of $\Delta V$ obtained from all transformations performed with a given $\epsilon$. Note that the error-bars on most data points are very small. 
    }
    \label{fig:exampleNNtest}
    \end{minipage}
\end{figure}

\paragraph{Symmetry detection}
After training, we use the neural network to predict $(\Delta V)_{\rm NN}$ \eqref{eq:DeltaV} for the validation data. 
In the presence of a symmetry, the converged neural network should predict $(\Delta V)_{\rm NN} \propto \epsilon^n $ with $n \geq 2 $ at leading order in $\epsilon$; in its absence, the leading term is $n =1 $. 
We can therefore deduce the presence of a symmetry from the absence of linear scaling for a large enough $\epsilon $-range of predictions.

The noise due to the neural network loss function 
is typically correlated with the magnitude of the input vectors $|\phi| $ and depends on details of the sampling.\footnote{This can in principle be utilized by only computing $(\Delta V)_{\rm NN} $ on data points $\phi$ for which the NN error $V_{\rm NN}(\phi) - V(\phi) $ is smaller than some tolerance $ \delta < \mape/100\%$.} As the magnitude of the transformation $\epsilon$ is chosen independently of the input data, the neural network noise is in principle uncorrelated with $\epsilon$. 
Then, the "error" in the prediction \eqref{eq:DeltaV} for a converged network is given by
\se{\label{eq:error}
\text{error} &=
(\Delta V)_{\rm NN} -(\Delta V)_{\rm truth} = \frac{V_{\rm NN}(\phi')-V(\phi')}{V(\phi)} \\
&\sim (1+\epsilon^n) \frac{  V_{\rm NN}(\phi)- V(\phi)}{V(\phi)}
+ \mathcal{O}(\epsilon^{n+1}) \\
&= (1+\epsilon^n) \frac{\mape}{100\%} + \mathcal{O}(\epsilon^{n+1})
}
We point out in particular that in the case of a symmetry transformation, no linear scaling is introduced in the error. 
Furthermore, we expect the scaling to become flat in $\epsilon$ for $\epsilon^n \lesssim \mape/100\%$.

We demonstrate the scaling of our converged network with a simple polynomial fit $ (\Delta V)_{\rm NN} = \sum_{i=0}^2 a_i \epsilon^i $ on a test set of 100  evenly spaced points in logspace in the interval $\epsilon = [10^{-3},1] $. For each value in $\epsilon$, we find $\Delta V$ by averaging over 2000 transformations where the original points in field space are randomly chosen from our validation set. We plot the the data points corresponding to the obtained values for $(\Delta V)_{\rm NN}$ and $(\Delta V)_{\rm truth}$ from SO(8) and SU(3) transformations in the NL$\Sigma$M in Fig.~\ref{fig:exampleNNtest}. We give the resulting fits for both models (NL$\Sigma$M and L$\Sigma$M) in Table~\ref{tab:fit}. 

Both Fig.~\ref{fig:exampleNNtest} and the fit coefficients in Table~\ref{tab:fit} demonstrate that there is a constant error in $(\Delta V)_{\rm NN}$ which approximately corresponds to the value of $\mape$ for the network. Even by eye one can identify the linear or quadratic scaling in $\Delta V$ from Fig.~\ref{fig:exampleNNtest}.  We find that we can correctly show that $a_1 \ll a_2$ for the SU(3) transformations in the NL$\Sigma$M, and for both SU(3) and SO(8) in the L$\Sigma$M model. We also correctly exclude $a_1 = 0$ for SO(8) in the NL$\Sigma$M, and for the arbitrary transformation \eqref{eq:faux} in the L$\Sigma$M.

\begin{figure*}
    \centering
    \includegraphics[width=0.45\textwidth]{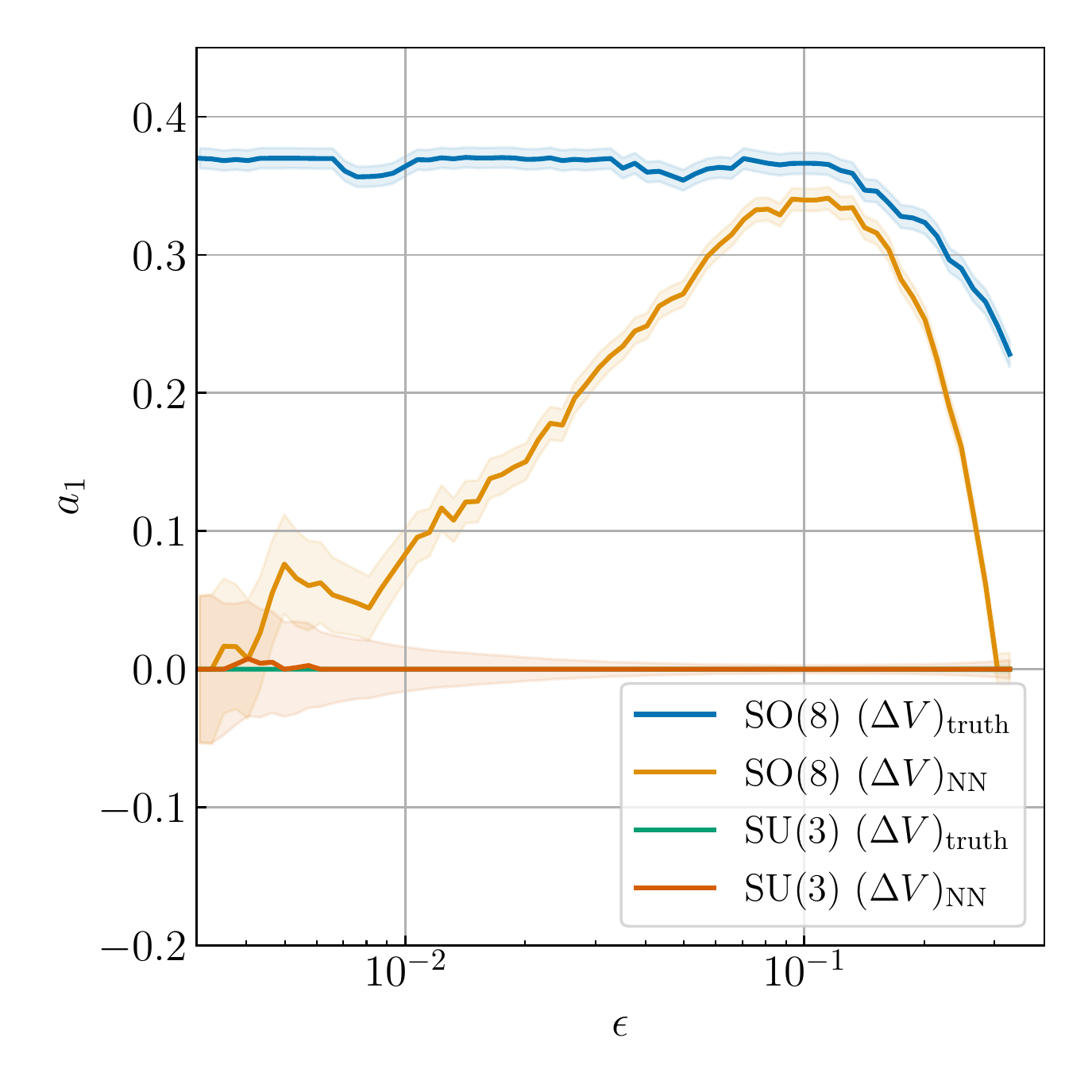}
    \includegraphics[width=.45\textwidth]{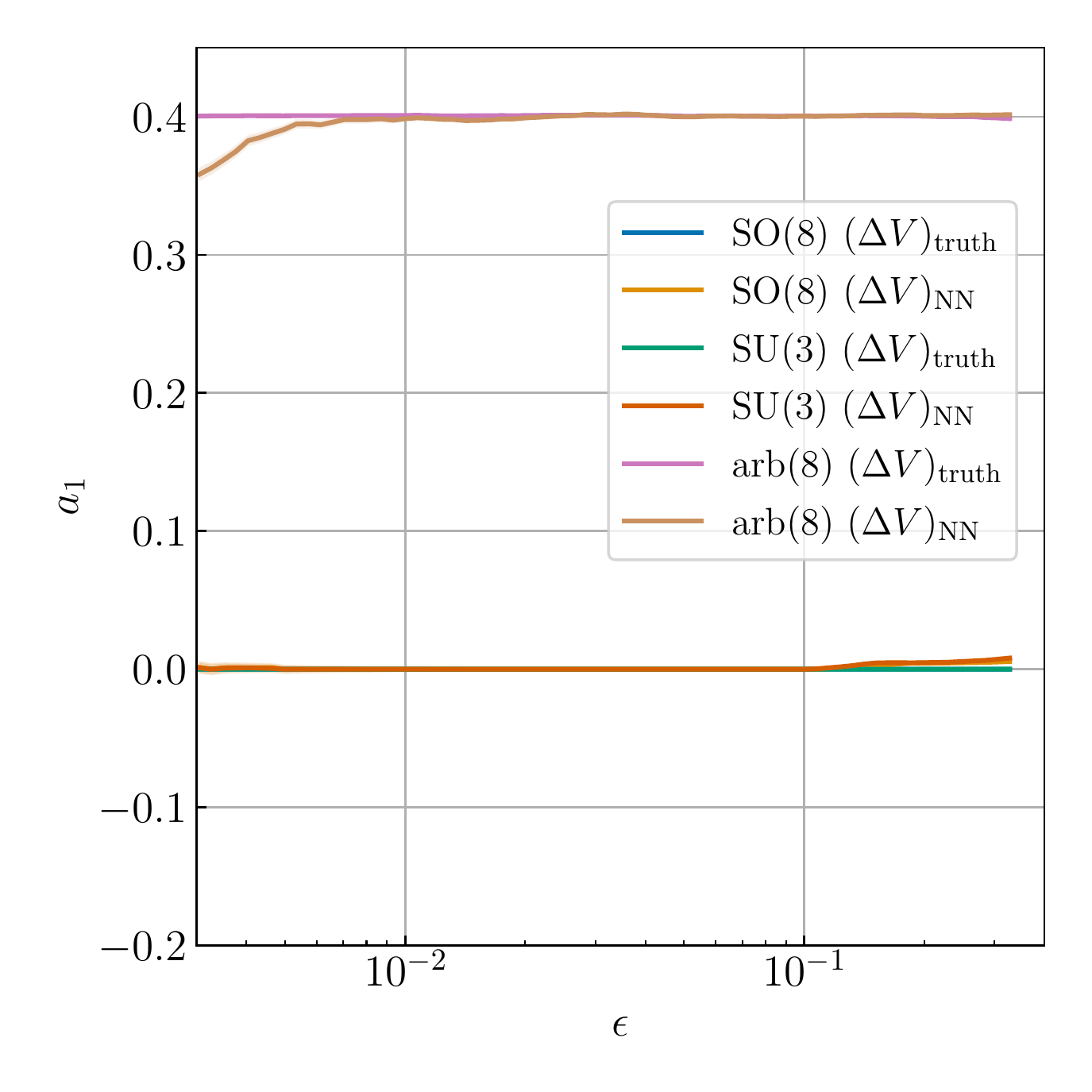}
    \caption{Value of the coefficient $a_1$ from the polynomial fit $ (\Delta V)_{\rm NN} = \sum_{i=0}^2 a_i \epsilon^i$ in the NL$\Sigma$M \eqref{eq:NLsigma} (left) and L$\Sigma$M \eqref{eq:Lsigma} (right) applied on a sliding window of $\epsilon$. The size of the window is one order of magnitude in $\epsilon$. We plot the value of $a_1$ against the median value of $\epsilon$ used in the fit. The shaded region corresponds to $\pm 1\sigma$ for the fit parameter. } \label{fig:slidingwindow}
\end{figure*}

We also check that linear scaling in $\epsilon$ doesn't appear for the $\epsilon$ range we consider. To  test this, we construct a sliding window in $\epsilon$ with a width corresponding to an order of magnitude in logspace. On this window we evaluate our simple polynomial fit on our data points for $\Delta V$. In Fig.~\ref{fig:slidingwindow} we plot the resulting values for $a_1$ as a function of $\epsilon$.  In both the NL$\Sigma$M and L$\Sigma$M we find that for transformations that preserve a symmetry, both $(\Delta V)_{\rm truth}$ and $(\Delta V)_{\rm NN}$ are consistent with $a_1=0$ for all $\epsilon \lesssim 1$, whereas for other transformations $a_1=0$ is excluded for a significant range of $\epsilon \ll 1$.

\paragraph{Results and discussion}
In this work we have proposed and demonstrated a method to detect a Lie group symmetry in a dataset using regression by an artificial NN. The NN was trained to replicate $V(\phi, \partial_\mu \phi)$ given training data in the form of $\{\phi, \partial_\mu \phi, V\}$. The symmetry was then tested by measuring the NN response to an $\mathcal{O}(\epsilon)$ transformation of the input fields according to the Lie algebra associated with the Lie group symmetry, effectively augmenting the dataset. We used this method to test for $SO(8)$ and $SU(3)$ symmetries in the  NL$\Sigma$M and L$\Sigma$M, see Fig.~\ref{fig:exampleNNtest}. As expected, we found the NL$\Sigma$M is symmetric under $SU(3)$ transformations, but is not invariant under $SO(8)$. For the L$\Sigma$M, we detected the presence of both $SU(3)$ and $SO(8)$ symmetries. 

\begin{table*}[hpbt]
\begin{center}
\begin{tabular}{|l | l || l | l | c !{\vrule width 1pt} c !{\vrule width 1pt} c | } 
 \cline{1-5} 
 \clineB{6-6}{2}
 \arrayrulecolor{black}
\cline{7-7}
 Model $\quad$  & $\bar{E}_{\%} $ &
 $\{T^a \}$ 
 & truth? &  $a_0$ & $a_1$ & $a_2$  \\ [0.5ex] 
 \hline\hline
 L$\Sigma$M & $0.05 \% $ & SO(8) & \checkmark  &  $(4.18 \pm 0.02) \times 10^{-4}$ & $(0.00 \pm 9.07) \times 10^{-5}$ & $0.390 \pm 0.001$   \\ 
 & & SU(3) & \checkmark &  $(4.14 \pm 0.02) \times 10^{-4}$ & $(0.00 \pm 1.13) \times 10^{-4}$ & $0.588 \pm 0.001$ \\ 
 & & arb(8)
 &  &  $(2.19 \pm 0.02) \times 10^{-4}$ & $0.4001 \pm 0.0001 $ & $(6.50 \pm 0.02)\times 10^{-2}$  \\ [0.5ex]
 \hline
 \hline
 NL$\Sigma$M & $ 0.7 \%$ & SO(8) & &  $(6.74 \pm 0.03) \times 10^{-3}$ & $0.215 \pm 0.002$ & $0.691 \pm 0.008$   \\ 
 & & SU(3) & \checkmark &  $(6.71 \pm 0.03) \times 10^{-3}$ & $(0.00 \pm 1.02) \times 10^{-3}$ & $0.758\pm 0.005$   \\ 
 \cline{1-5} 
 \clineB{6-6}{2}
 \arrayrulecolor{black}
\cline{7-7}
\end{tabular}
\caption{\small Polynomial fit $ (\Delta V)_{\rm NN} = \sum_{i=0}^2 a_i \epsilon^i $ over the full interval $\epsilon = [10^{-3},1] $ quoted with $1 \sigma $ error bars. It is seen that $ a_0 \sim \mape/100\%$ and $a_1$ is nonzero in the absence of symmetry. 
}
\label{tab:fit}
\end{center}
\end{table*}

The method presented here takes advantage of the fact that the Lie algebra lives in the tangent space of the group's manifold. This mitigates the importance of perfect interpolation as well as exact invariance under the full symmetry group: a symmetric system's true potential will not be exactly invariant under the Lie algebra transformation, but instead will exhibit $\mathcal{O}(\epsilon^2)$ scaling.
In contrast, a system that lacks the symmetry will instead exhibit linear $\mathcal{O}(\epsilon)$ scaling. By ruling out $\mathcal{O}(\epsilon)$ scaling, we can rule out the \it absence \rm of a symmetry.

The power of the neural network lies in the ability to extend this method to more realistic scenarios in which the symmetry is obscured. The next steps are to apply this technique to recover the same symmetry from more realistic data limited by minimal experimental signals or contaminated by noise.
Data from more realistic experimental signals would not in general provide an ordering for the NGB fields. The generators of $SU(N)$ do not commute with the operator that shuffles these fields, and so this method would only recover the symmetry in one of the $(N^2-1)!$ combinations of shuffled NGB fields.\footnote{We note that the possibility of indistinguishable pNGB fields is not particularly worrisome. The assumption that these fields are massive implies the presence of explicit symmetry breaking of the SSB group, generally allowing the NGB fields to be distinguished.
This the case for pions and kaons transforming under the $SU(3)$ flavor symmetry of QCD and played a key role in the discovery of the 8-fold way.}
For $SU(3)$, we are able to reorder the shuffled fields by exploiting properties of members of the Cartan subalgebra, $T_3$ and $T_8$. This trick may be formalized and extended to general $SU(N)$ or even general Lie Groups, but we leave this for future study. 
In future work we will also study the use of this method
to recover approximate symmetries in the presence of explicit breaking.

\vspace{5mm}
{\bf Acknowledgments --} 
The authors thank Steve Abel, Ben Allanach, Juan C. Criado, Jeff Dror, Sven Krippendorf, Dalimil Maz\'{a}\u{c}, and Michael Spannowsky for useful discussions.
D. Croon thanks the Aspen Center for Physics (supported by NSF grant PHY-1607611) and the Galileo Galilei Institute
for hospitality during the completion of this paper. D. Cutting is supported by the Academy of Finland grants 328958 and 345070.
D. Croon and R. Houtz are supported by the STFC under Grant No. ST/T001011/1.

\bibliographystyle{apsrev4-1}
\bibliography{refs}

\clearpage

\onecolumngrid
\section{Supplementary material}
\label{app:NN}
\paragraph{Neural network optimization}
Before the analysis in the main paper was carried out,  a number of optimization studies were performed which we will describe here.
Initial studies included linear regression and decision tree methods. Ultimately for the models discussed in this work, better convergence was found with the discussed sequential feed-forward neural networks.

In our optimization studies, we 
fixed all but one -- or in some cases two -- hyperparameters. This optimization was sufficient to produce networks, performing at the sub percentage level validation mean absolute percentage error ($\mape$). The hyperparameters in Table~\ref{tab:nnhypers} were used as a default from which to improve on, and the range of hyperparameter testing is given in Table~\ref{tab:nn-hypers-test}. Network performance was assessed with the minimum value of the  $\mape$. The minimum  $\mape$ was recorded for ten networks of each configuration, and the corresponding mean and standard error were calculated. 

The optimal shape of the network was found at seven hidden layers with $410\pm 25$ neurons per layer. However, the variance of the network performance effectively plateaued, and the stochastic nature of network training dominated the variation in the $ \mape$ for networks with more than three hidden layers and two hundred neurons. Typically, increased network size increases the danger of over-fitting. Over fitting was only seen with small numbers of input training data points. Training time and forward propagation time can be reduced by using smaller networks. Thus, a network with three hidden layers and two hundred neurons per layer is advised. However, when using more training data points, one can see performance improvements from increased degrees of freedom. 

The Adam optimizer performs very well across a wide variety of tasks. This problem was no exception, with Adam providing modest improvements over Nadam and Adamax. Notably, SGD failed to converge, implying that per-parameter historical weighting was essential. In addition, the use of a learning rate reduction after a set number of epochs was seen to improve  $\mape$ with large data-sets; however, this was not  tested rigorously. Adjustment of the $\beta_1$ and $\beta_2$ hyperparameters was investigated,  though network performance was agnostic to both parameters with values above 0.8. Although the Adam optimizer possesses an adaptive learning rate, extreme learning rate values yielded the expected poor convergence. A learning rate in the region of $[4\cdot 10^{-4},2 \cdot 10^{-2}]$ provided constant  $\mape$.

Network performance should scale with training data size, and this was seen. The network performance exhibited power-law scaling with training data size, and exponentially more data points are required to improve performance by the same amount.  Thus, training time becomes the primary constraint on performance improvement. 
The variation of network performance with batch size was negatively correlated. The optimal performance of the network was found with mini-batches of size 8. Below this, network, convergence was unreliable. The preference of the network to lower batch sizes is in line with other typical training results. Training time is drastically affected by lowering the batch size as parallelization is reduced. However, the performance improvements are significant enough to suggest a batch size of 8.

\begin{table*}[b]
\begin{minipage}{.75\textwidth}
\begin{tabular}{|l | l |l |}
 \hline
 Hyperparameter $\quad$ & Default Value $\quad$ & Test Range \\ [0.5ex] 
 \hline\hline
 Hidden layers* & 3 &$[2,8]$\\ 
 \hline
 Neurons/layer* & 512 & $[10,510]$ \\
 \hline
 Optimizer & Adam &   SGD, RMSprop, Adam,
 \\ 
 & & NAdam, Adamax\\
 \hline
 Adam: $\beta_1$ and $\beta_2 $ & $\beta_1 = 0.9, \, \beta_2 = 0.99 \quad$ &$\beta_1  [0.5,0.99], \, \beta_2 [0.5,0.99]$\\
 \hline 
 Learning Rate & 0.001 &$[10^{-4},10^{-1}]$ \\
 \hline
 Loss function & MAPE & MAPE, MAE, MSE, 
 \\ 
 & & Huber Loss\\
 \hline
 Training epochs & 100 & $[0,100]$ \\
 \hline
 Training set size & $10^5$ &$[16,10^5]$\\
 \hline
 Batch size & 256& $[4,500]$ \\
 \hline
\end{tabular}
\caption{\small 
The hyperparameters explored while analyzing network performance.
*Hidden layers and neuron count were varied simultaneously}
\label{tab:nn-hypers-test}
\end{minipage}
\end{table*}
\twocolumngrid

\end{document}